# Eu-doped CsSrCl$_3$ Large Nanocrystal Clusters with Self-Reduction Effect and Near-Unity Quantum Yield

Chuangchang Lei[1], Xiang Wu[1], Yaohua Li[1], Xu Xu[2], Guangzheng Zuo[1], Qiongrong Ou[1], Shuyu Zhang[1,*]

[1] State Key Laboratory of Photovoltaic Science & Technology, School of Intelligent Robotics and Advanced Manufacturing, Institute for Electric Light Sources, Fudan University, Shanghai 200433, PR China

[2] Donghai Laboratory, Zhoushan, 316021, PR China

E-mail: forxuxu@126.com; gzzuo@fudan.edu.cn

*Corresponding author: zhangshuyu@fudan.edu.cn

# Abstract

Europium halide perovskites have emerged as promising candidates for environmental-friendly blue-emitting materials. However, their development is hindered by relative low photoluminescence quantum yields (PLQY, e.g. ~2-5% for intrinsic $CsEuCl_3$) and poor stability against air. Here, we introduce a one-step-procedure for synthesizing $Eu^{2+}$-doped $CsSrCl_3$ large nanocrystal clusters (LNCs) with the effect of self-reduction, therefore eliminating the use of conventional reductant oleylamine (OAm) and ensuring phase purity. The $CsSrCl_3$:Eu LNCs shows photoluminescence emission centered at 430 nm with a full width at half-maximum (FWHM) of 25 nm and a PLQY of ~40%, which can be further enhanced to ~97% after passivating the surface defects by adding trioctylphosphine (TOP), the highest among all reported lead-free blue-emitting perovskite nanocrystals. The stability of $CsSrCl_3$:Eu can also be improved significantly by epitaxially growing ZnS shell on the surface. This work will shed more light on lanthanide and alkaline-earth metal (AEM)-based perovskites for nontoxic light-emitting materials.

# Introduction

Lead-based halide perovskite nanocrystals have become a promising class of materials for various optoelectronic devices due to their superior optical properties[1–5]. However, the inherent toxicity of lead ions poses a major challenge to its commercialization. Although numerous researchers have developed lead-free double perovskites ($Cs_2AgBiBr_6$, $Cs_2AgInCl_6$, etc.)[6,7] and vacancy-ordered perovskites ($Cs_3Bi_2Br_9$, $Cs_3Sb_2Br_9$, etc.)[8,9] with high PLQY, they still suffer from a broad spectral widths of 40-200 nm originating from self-trapped excitons. Europium-based halide perovskites $CsEuX_3$ (X = Cl, Br, I) have been considered promising candidates as lead-free perovskites due to their potential of intense photoluminescence with narrow emission line widths. Given the similar ion radii of $Eu^{2+}$ and $Pb^{2+}$ (117 pm for $Eu^{2+}$ vs 119 pm for $Pb^{2+}$), $Eu^{2+}$ ions are able to completely substitute $Pb^{2+}$ with an appropriate Goldschmidt tolerance factor[10,11]. Furthermore, attributing to the dipole-allowed 4f-5d electronic transitions, $CsEuX_3$ exhibits a narrow emission line width (<30 nm) in the deep-blue spectral region[12–17].

However, there still exist some critical challenges to obtain phase-pure europium-based halide perovskites with both high PLQY and acceptable air stability. On one hand, the conventional approach to prepare a soluble Eu(II) precursor is to reduce a trivalent Eu(III) precursor with a reducing agent oleylamine (OAm) and get the Eu(II)-OAm complex. Nonetheless, it is observed that excess OAm will inevitably lead to the formation of weakly emissive impurities like $CsBr:Eu^{2+}$ instead of perovskites, which is attributed to the strong affinity of $Eu^{2+}$ to OAm[12,14]. On the other hand, the PLQY of CsEuX$_3$ is hindered by both the surface and internal defects. CsEuCl$_3$ nanocrystals exhibited a low intrinsic PLQY of ~2.0%, which could be increased to ~5.7% via post-synthetic surface treatment[12] or to ~19.7% via Ni$^{2+}$ ion doping[15]. To avoid the defects possibly caused by the Cl element, CsEuBr$_3$ nanocrystals were synthesized and exhibited a higher intrinsic PLQY of 52%, the value of which could be elevated to 70% after surface passivation[13]. Meanwhile, another work also confirmed a PLQY of 40.5% for CsEuBr$_3$ nanocrystals synthesized with a different bromide precursor[14]. Despite the important work done to optimize phase purity and PLQY, all europium-based halide perovskites still suffer from poor stability in air due to the high reactivity of Eu$^{2+}$ with moisture and oxygen, which will lead to the formation of Eu$^{3+}$ and decomposition of nanocrystals.

Here, we demonstrate a facile solution-phase synthesis method of Eu$^{2+}$-doped strontium-based halide perovskite CsSrCl$_3$:Eu LNCs, realizing the reduction of Eu$^{3+}$ into Eu$^{2+}$ by charge compensation mechanism in AEM-based halide perovskites, therefore eliminating the use of OAm and ensuring phase purity. A descriptor based on host structure stability and electronegativity difference between AEM and halogens is proposed to evaluate the effectiveness of self-reduction, and CsSrCl$_3$ is chosen due to its strongest self-reduction effect. Considering that Sr$^{2+}$ has almost the same radius with Eu$^{2+}$ (118 pm for Sr$^{2+}$ vs 117 pm for Eu$^{2+}$), it is also promising to synthesize CsSrCl$_3$ with a satisfying Goldschmidt tolerance factor. Notably, CsSrCl$_3$:Eu LNCs exhibits an intrinsic PLQY of ~40% and a full width at half-maximum of 25 nm at 430 nm. The value of PLQY can be further increased to 97% by adding trioctylphosphine (TOP) as a post-synthetic surface treatment procedure. After epitaxially growing a ZnS shell on the surface of LNCs, the stability of CsSrCl$_3$:Eu in ambient environment can also be impressively improved. This work provides a new perspective for the development of lead-free halide perovskites and offers a promising blue-lighting perovskite with both high PLQY and remarkable stability.

## Results

### 1. Growth kinetics and self-reduction effect of CsSrCl$_3$:Eu

The one-step-synthesis of CsSrCl$_3$:Eu is carried out by injecting benzoyl chloride (Bz-Cl) into the precursors within three-necked flask at 290°C, as shown in Supplementary Fig. 1. After 10 min, the reaction was quenched using an ice-water bath (see Methods for details). To gain a comprehensive understanding of the growth kinetics of CsSrCl$_3$:Eu, we systematically analyzed the samples obtained at various durations of reaction, from 1 s to 10 min. All samples match well with the standard pattern of CsSrCl$_3$ (*P4mm*, PDF no. 00-020-0289, $a = b = 5.59$ Å, $c = 5.62$ Å) (Fig. 1a and Supplementary Fig. 2). It is noticed that CsSrCl$_3$ emerges even the reaction lasted for only 1 s (LNCs-1s), and there can hardly be found any other phases belonging to possible impurities. As the reaction goes on, the XRD peaks of CsSrCl$_3$ increase gradually and become more distinguishable after 10 min (LNCs-10min), indicating the improved crystallinity of samples. Compared to the time-consuming synthesis of conventional europium-based perovskites (~60 min)[12–15], the easy and fast formation of CsSrCl$_3$ host structure is mainly attributed to the weak binding between Sr$^{2+}$ and oleic acid (OA) (binding energy of -147.06 kcal/mol, see Supplementary Note 1), thus fundamentally preventing any possible impurities. In comparison, the strong binding of $Eu^{2+}$ with OAm (binding energy of -538.76 kcal/mol) not only makes it difficult for Eu(II)-OAm to enter into the host structure due to its low reactivity, but also increases the steric hindrance, therefore massively slows down the reaction rate and leads to the formation of impurities inevitably[14].

The fast formation and growth kinetics of $CsSrCl_3$ host structure can also be verified by the transmission electron microscopy (TEM) images. As can be seen in Fig. 1b, the particles of LNCs-1s range randomly from ~100 nm to ~2 μm, larger than most reported $CsEuX_3$ nanocrystals (<50 nm)[12–16], and are clearly separated from each other. After 10 min of reaction, the interfaces disappear and a large nanocrystal cluster with an irregular shape is formed eventually in LNCs-10min (Fig. 1c). Notably, the high-resolution TEM (HRTEM) images of LNCs-10min (Fig. 1d-g) exhibit an orderly arrangement spacing of 0.323 nm, which corresponds to the (1 1 1) crystal facets of $CsSrCl_3$. The continuous growth of crystal facets in a specific direction over a large scale (>100 nm) further confirms the improved crystallinity and rapid growth characteristics of the host structure. The relatively large size of LNCs and their irregular shapes are mainly attributed to the high reactivity of $Sr^{2+}$-OA and the highly dynamic binding feature of OA to the LNCs surface[18], leading to continuous filling of free $Sr^{2+}$ ions onto the surface. As a result, the interfaces are gradually filled up with free ions outside the LNCs, serving as an expansion of the initial structure. The relative ratio of elements obtained from inductively coupled plasma mass spectrometry (ICP-MS) also shows that the proportion of Sr in LNCs-10min is much higher than that in LNCs-1s (Supplementary Fig. 3), confirming the incorporation of more $Sr^{2+}$ ions into the host structure as the reaction time increases. In addition, it is also worth mentioning that the ratio of Sr:Cs in LNCs-10min is 1.25:1, higher than the ideal 1:1 in $CsSrCl_3$, indicating a Sr-rich surface for LNCs-10min.

According to energy-dispersive spectroscopy (EDS) mapping of LNCs-1s, Cs, Sr, Cl and Eu are all uniformly distributed in the LNCs (Supplementary Fig. 4), demonstrating the successful doping of Eu. Additionally, X-ray photoelectron spectroscopy (XPS) analysis of LNCs-10min detected intense signals of the $3d_{3/2}$ and $3d_{5/2}$ peaks of $Eu^{2+}$ (Fig. 2a), proving the occurrence of self-reduction in strontium-based perovskite. The presence of $3d_{3/2}$ and $3d_{5/2}$ peaks of $Eu^{3+}$ mainly comes from the inevitable oxidation of $Eu^{2+}$ during measurement, owing to the high reactivity of $Eu^{2+}$ with moisture and oxygen in air. Here, LNCs with different reaction time all exhibit an intense blue emission positioned at around 430 nm with FHWM less than 30 nm (Fig. 2b, Supplementary Fig. 5 for the full range from 400-700 nm), which is the characteristic photoluminescence spectrum of $[EuCl_6]^{4-}$ in phase-pure products[12,15,16], thereby indicating the successful replacement of $Sr^{2+}$ by $Eu^{2+}$ and supporting the reduction of $Eu^{3+}$ to $Eu^{2+}$. Furthermore, as the reaction goes on, FWHM of LNCs narrows from 26.1 nm to 24.2 nm (inset of Fig. 2b), originating from the improved crystallinity as mentioned before.

The self-reduction of $Eu^{3+}$ to $Eu^{2+}$ is attributed to the charge compensation mechanism[19,20]. Under the law of charge conservation, the whole process can be expressed by the following Eqs. 1-3:

$$3Sr^{2+} \xrightarrow{2Eu^{3+}} V_{Sr}^{\times\times} + 2Eu_{Sr}^{\bullet} \tag{1}$$

$$V_{Sr}^{\times\times} \rightarrow V_{Sr} + 2e \tag{2}$$

$$2Eu_{Sr}^{\bullet} + 2e \rightarrow 2Eu_{Sr} \tag{3}$$

where $Eu_{Sr}^{\bullet}$ represents a single-positively-charged $Eu^{3+}$ site that replaces the $Sr^{2+}$ site and $V_{Sr}^{\times\times}$ represents the generated vacancy with two negative charges. As to the remaining $V_{Sr}$ after charge compensation, it will highly probably be filled with another $Sr^{2+}$ in the structure, since there are obviously more Sr detected from the ICP analysis in LNCs-10min.

The strength of self-reduction in AEM-based halide perovskites is significantly influenced by the electronegativity difference ($\triangle EN$) between AEM and halogens and the structural stability of host structure, and we experimentally determined the reduction degree based on the ratio of PL intensity of $I(Eu^{2+})$ at ~430 nm to $I(Eu^{3+})$ at 614 nm. As can be seen in Fig. 2c, all samples exhibit the characteristic peak of $[EuX_6]^{4-}$ at ~430 nm, which indicates the successful reduction of $Eu^{3+}$ to $Eu^{2+}$. Among all these successfully synthesized Eu-doped AEM-based halide perovskites, $CsSrCl_3$:Eu shows negligible $Eu^{3+}$ emission, while $CsCaBr_3$:Eu shows the strongest. On one hand, after comparing $I(Eu^{2+})/I(Eu^{3+})$ with $\triangle EN$ (Supplementary Fig. 6a), it is easy to find that the larger $\triangle EN$ is, the stronger the self-reduction effect will be. Since as $\triangle EN$ increases, less charge will be concentrated on AEM cations and vacancies[21], which therefore makes it easier for negative charges to be transferred to $Eu^{3+}$. On the other hand, the structural stability of $ABX_3$ halide perovskites can be estimated by both Goldschmidt tolerance factor ($t$) and octahedral factor ($\mu$), and it has been empirically concluded that the formation of $ABX_3$ halide perovskites requires $0.81 < t < 1.11$ and $0.44 < \mu < 0.90$[22]. After plotting $I(Eu^{2+})/I(Eu^{3+})$ against $(\mu+t)$ (Supplementary Fig. 6b), it can be found that as $(\mu+t)$ increases, the ratio of $I(Eu^{2+})/I(Eu^{3+})$ increases as well, indicating a stronger self-reduction effect within a host structure with better structural stability. A host structure with better stability is less likely to decompose when $Eu^{3+}$ ions enter and generate vacancies, as a result, more vacancies are able to be retained and propel self-reduction, leading to a higher ratio of $I(Eu^{2+})/I(Eu^{3+})$. Consequently, the effectiveness of self-reduction in AEM-based halide perovskites is determined simultaneously by host structure stability $(\mu+t)$ and $\triangle EN$, which can be expressed as $\triangle EN^{(\mu+t)}$, and the larger the descriptor is, the stronger the self-reduction effect will be, thus a higher ratio of $I(Eu^{2+})/I(Eu^{3+})$ (Fig. 2d). After exponentially fitting $I(Eu^{2+})/I(Eu^{3+})$ with the variables mentioned above, $\triangle EN^{(\mu+t)}$ shows a determination coefficient ($R^2$) of 0.9996, proving its effectiveness of evaluating the strength of self-reduction. The values of $t$, $\mu$ and $\triangle EN$ for different AEM-based perovskites can be found in Supplementary Table 1.

## 2. Improving the optical properties with the post-synthetic treatment of TOP

The PLQY of LNCs rises from ~7% to ~40% as the reaction time increases during the initial 10 min and stays at ~40% even the reaction time is extended to 30 min (Pristine-LNCs) (Supplementary Fig. 7). Compared to previously reported $CsEuCl_3$[12,15], $CsSrCl_3$:Eu demonstrates a much higher intrinsic PLQY, which is mainly attributed to the improved crystallinity. The high reactivity of Sr-OA precursor drives more $Sr^{2+}$ to actively occupy vacancy defects both in the interior and on the surface of LNCs, resulting in a Sr-rich structure environment (as proved above) and therefore significantly reducing the number of non-radiative recombination centers. To further increase the PLQY of LNCs, we propose a post-synthetic surface treatment by directly injecting appropriate amount of TOP into the reaction system ten minutes after the start of reaction. As can be seen from XRD in Supplementary Fig. 8, the obtained TOP-LNCs is still a phase-pure product and has an even better crystallinity compared to Pristine-LNCs. Notably, TOP-LNCs achieves near-unity PLQY (97%) while maintaining a narrow emission line width (FWHM = 25 nm) at 430 nm (Fig. 3a). The photoluminescence excitation (PLE) spectra were conducted at 430 nm emission wavelength, and both Pristine-LNCs and TOP-LNCs exhibit strong excitonic absorption at ~350 nm.

Time-resolved photoluminescence (TRPL) in Supplementary Fig. 9 shows the PL decay curves of Pristine- and TOP-LNCs, both of which can be fitted with biexponential decay functions. Pristine- and TOP-LNCs exhibit similar average PL lifetimes, 836 ns and 854 ns, respectively, much longer than that of reported $CsEuCl_3$ (≤30 ns)[12,15–17]. The value of radiative recombination ($k_r$) increased from 0.514 $\mu s^{-1}$ to 1.113 $\mu s^{-1}$ after the treatment of TOP, and nonradiative recombination ($k_{nr}$) dropped from 0.682 $\mu s^{-1}$ to 0.035 $\mu s^{-1}$ (detailed in Supplementary Note 2 and Supplementary Table 2). This pronounced variation indicates that the promoted PLQY (from ~40% to 97%) mainly arises from the passivation of nonradiative recombination defects on the surface of LNCs with the assistance of TOP. Additionally, the exciton binding energy can be estimated by temperature-dependent PL spectra in Supplementary Fig. 10 (detailed in Supplementary Note 3). TOP-LNCs exhibits an exciton binding energy ($E_b$) of 73.03 meV (Supplementary Fig. 11), larger than that of most 3D and bulk lead-based halide nanocrystals (<50 meV)[2,23,24], which allows the formation of stable excitons and contributing to the radiative recombination. Furthermore, we also calculated its Huang-Rhys factor $S$ and $\hbar\omega_{phonon}$ (detailed in Supplementary Note 3). As depicted in Supplementary Fig. 11, $S$ = 2.18 and $\hbar\omega_{phonon}$ = 36.44 meV, which is very similar to those in $CsPbBr_3$ ($S$ = 3.223 and $\hbar\omega_{phonon}$ = 28.011 meV)[25], indicating a weak electron-phonon coupling strength and therefore a good monochromaticity in TOP-LNCs.

The enhancement of optical properties originating from surface passivation is further verified by quantitative X-ray photoelectron spectroscopy (XPS) analysis. Fig. 3b displayed an increasing trend in the relative ratio of Cs:(Sr+Eu) from 0.53:1.00 to 0.81:1.00 after the treatment of TOP, indicating the adsorption of extra $Cs^+$ ions on the surface of LNCs. Besides, an increase of Cl:(Sr+Eu) was also detected, rising from 2.28:1.00 to 2.92:1.00. Apparently, to maintain the charge conservation on the surface of LNCs, free $Cl^-$ ions in the reaction system will also come to coordinate with part of these extra $Cs^+$ ions and incidentally passivate some Cl vacancies on the surface. By comparing the XPS results of Pristine-LNCs with TOP-LNCs (Fig. 3c-e), it can be observed that, after the surface treatment of TOP, the core-level peaks of Cs $3d_{3/2}$ and Cs $3d_{5/2}$ both shift toward higher binding energies by 0.4 eV and 0.5eV respectively. Meanwhile, there is hardly any change to the core-level peaks of Sr and Cl. Such difference demonstrates an increase of cationic charge on $Cs^+$, which comes from the extra $Cs^+$ adsorbed on the surface of LNCs with the assistance of TOP. Since these new ions are not fully coordinated with $Cl^-$, there will be less electron donors around them, thus shifting its binding energies towards higher values.

Fourier transform infrared spectroscopy (FTIR) was conducted to confirm the binding of TOP on the surface of LNCs (Fig. 3f). It is shown that the spectra of Pristine- and TOP-LNCs are mostly similar with each other, while a new signal located at 1136 $cm^{-1}$ emerges in TOP-LNCs. This new peak is consistent with the characteristic C-P stretching mode in TOP between 1049 $cm^{-1}$ to 1213 $cm^{-1}$, indicating that TOP is attached to the surface of LNCs[26–28]. Moreover, the position of this new peak was different from the peaks in pure TOP, also revealing the interaction between TOP and LNCs. The mechanism of surface treatment by TOP is summarized as Fig. 3g.

### 3. Enhancing air stability by coating ZnS shell

The biggest challenge for europium-based halide perovskites lies in its extremely high sensitivity to moisture and oxygen, the emission of thin films made from which will disappear within several seconds as soon as they are exposed to the air[12,15,16]. In the case of $CsSrCl_3$:Eu, we investigated the optical stability of Pristine- and TOP-LNCs in the ambient environment (~25°C, ~40% humidity) under continuous irradiation from a 365 nm UV lamp with a power density of 20 $mW/cm^2$. As shown in Fig. 4a, the PL intensity of Pristine-LNCs remained ~78% of its original value even after 10 mins of irradiation. However, the peak position shifted to 425 nm after 35 mins, suggesting the decomposition of coordination environment around $Eu^{2+}$. With regard to TOP-LNCs in Fig. 4b, it dropped to ~66% of its original intensity after 10 mins, while the peak position didn't shift even after 40 mins. The stability performance of both Pristine- and TOP-LNCs suggests that the intrinsic stability of strontium-based halide perovskite is much higher than europium-based halide perovskite, since the Sr-rich surface protects most Eu from direct contact with moisture and oxygen, therefore considerably decelerating the decomposition of perovskite structure. Additionally, the treatment of TOP is of great help for the rigidity of perovskite structure, maintaining the structural environment around $Eu^{2+}$ in 40 min.

To further enhance the stability of LNCs against moisture, oxygen and continuous UV irradiation, we proposed another post-synthetic treatment strategy, aimed at growing a ZnS shell on the surface of LNCs (ZnS-LNCs) (see Methods for details). Considering the fact that the lattice mismatch between (2 0 0) plane of $CsSrCl_3$ (0.279 nm) and (2 0 0) plane of ZnS (0.271 nm) is only ~3% (less than 15%)[29,30], it is very promising to realize the epitaxial growth of ZnS[31]. Based on the EDS image of ZnS-LNCs in Supplementary Fig. 12, Zn and S elements are uniformly distributed on the surface of LNCs. As we can see from the TEM image in Supplementary Fig. 13, the interplanar spacing of 0.270 nm belongs to the (2 0 0) spacing of ZnS. In addition, after the growth of ZnS, the PLQY of LNCs reaches 90% (Supplementary Fig. 14), which is mainly attributed to the passivation from $Zn^{2+}$ with a small radius of 74 pm and excess $Cl^-$ onto the surface. Subsequently, we assessed the stability of ZnS-LNCs under the same condition with Pristine-LNCs and TOP-LNCs (Fig. 4c). It is observed that the center of PL didn't change even after 7 hours of continuous irradiation. To better illustrate the difference, we plotted the curves of their maximum PL intensity as a function of time, as shown in Fig. 4d. Notably, the time needed for the intensity of ZnS-LNCs to drop to 70% of its initial intensity ($T_{70}$) is ~85 min, which is more than 5 times longer than that of Pristine-LNCs (~15 min), demonstrating a significant improvement in the stability of LNCs after the growth of ZnS shell. Additionally, the water contact angle (CA) was measured to verify the moisture resistance of ZnS-LNCs (Supplementary Fig. 15). Compared to Pristine-LNCs (~44°), the CA of ZnS-LNCs increased to ~56°, indicating a better resistance against moisture.

## Discussion

In summary, we demonstrated the solution-phase synthesis of $Eu^{2+}$-doped strontium-based halide perovskite $CsSrCl_3$:Eu by means of self-reduction. The charge compensation mechanism ensured the successful reduction of $Eu^{3+}$ to $Eu^{2+}$ without the use of reductant ligands, and therefore led to the growth of large nanocrystal clusters and improved crystallinity. Optical measurements revealed that $CsSrCl_3$:Eu exhibits a narrow FWHM of 25 nm at 430 nm with an intrinsic PLQY of ~40%. Furthermore, the value of PLQY can be enhanced to 97% by using TOP as a post-synthetic treatment procedure to passivate the surface vacancies. By epitaxially growing a shell of ZnS on the surface, the stability of $CsSrCl_3$:Eu against moisture, oxygen and continuous irradiation improved significantly. This work introduces a new perspective for the development of lead-free perovskites by offering a Eu-doped strontium-based perovskite with near-unity PLQY, high monochromaticity and stability, demonstrating great potential for future applications including light-emitting diodes, photodetectors, fluorescent sensing and display technologies.

# Methods

**Materials**

Cesium carbonate ($Cs_2CO_3$, 99.9%), strontium oxide (SrO, 99.9%), europium acetate hydrate ($Eu(CH_3COO)_3·xH_2O$, 99.99%), calcium oxide (CaO, 99.9%), 1-octadecene (ODE, $C_{18}H_{36}$, 90%), oleic acid (OA, $C_{18}H_{34}O_2$, analytical reagent grade), oleylamine (OAm, $C_{18}H_{37}N$, 80-90%), benzoyl chloride (Bz-Cl, $C_7H_5ClO$, 99%), benzoyl bromide (Bz-Br, $C_7H_5BrO$, 98%), trioctylphosphine (TOP, $C_{24}H_{51}P$, ≥90%) and methyl acetate (MeOAc, $CH_3COOCH_3$, 99%) were purchased from Aladdin. N-hexene ($C_6H_{14}$, analytical reagent grade) was purchased from Sinopharm. Unless specifically noted, all chemicals were used without further purification.

**Preparation of a stock of sulfur-octadecene (S-ODE)**

All the chemicals were treated under an inert atmosphere in the glovebox. For the preparation of S-ODE solution, 32.07 mg of sulfur powder (1 mmol) and 5 mL ODE were transferred into a 25 mL three-necked flask and maintained under vacuum and continuous stirring for 30 min at 120°C to eliminate the oxygen. Thereafter, the temperature was raised to 190°C for 90 min under $N_2$ atmosphere. The obtained

S-ODE was cooled down to room temperature and then transferred to the glove box for further use.

**Synthesis of CsSrCl$_3$:Eu LNCs**

Prior to the synthesis of LNCs, europium acetate hydrate was dehydrated in a vacuum oven at 120°C for 12 hours to obtain anhydrous europium acetate. In a typical synthesis, 32.6 mg of Cs$_2$CO$_3$ (0.1 mmol), 21.8 mg of SrO (0.21 mmol), 29.6 mg of anhydrous Eu(CH$_3$COO)$_3$ (0.09 mmol), 2 mL OA and 5 mL ODE were loaded into a 25 mL three-necked flask and degassed under vacuum for 30 min at 120°C. After the dissolution of the mixture, the temperature was raised to 290°C under N$_2$. Subsequently, 0.15 mL Bz-Cl (1.29 mmol) was swiftly injected into the solution, and upon the completion of the reaction, the solution was quickly quenched to room temperature by using an ice-water bath. For the purification of CsSrCl$_3$:Eu LNCs, the crude solution was centrifuged at 9000 rpm for 5 min. The supernatant was discarded, while 3 mL of hexane and 3 mL of MeOAc were added to the precipitates and centrifuged again at 7000 rpm for 5 min. The final product was redispersed in 3 mL of hexane for further tests.

**Synthesis of CsSrBr$_3$:Eu, CsCaCl$_3$:Eu and CsCaBr$_3$:Eu**

CsSrBr$_3$:Eu, CsCaCl$_3$:Eu and CsCaBr$_3$:Eu were prepared following a procedure similar to that used for synthesizing CsSrCl$_3$:Eu. Except that 0.15 mL Bz-Br (1.26 mmol) was injected into the solution to form CsSrBr$_3$:Eu. 11.8 mg of CaO (0.21 mmol) was used for CsCaCl$_3$:Eu and 0.15 mL Bz-Cl (1.29 mmol) was injected into to solution. 11.8 mg of CaO (0.21 mmol) was used for CsCaBr$_3$:Eu and 0.15 mL Bz-Br (1.26 mmol) was injected into the solution. The purification is the same with the section above.

**Synthesis of TOP-LNCs**

After completing the growth of CsSrCl$_3$:Eu at 290°C as discussed in the section above, 0.2 mL TOP (0.45 mmol) was directly injected into the solution. The reaction temperature was maintained at 290°C for 10 min, and the solution was quenched to room temperature by using an ice-water bath. The purification is the same with the section above.

**Synthesis of ZnS-LNCs**

After completing the growth and centrifugation of CsSrCl$_3$:Eu, the final product was redispersed in 3 mL of ODE to serve as seed solution. For the synthesis of ZnS-LNCs, 27.3 mg ZnCl$_2$ (0.2 mmol), 0.1 mL OAm and 4 mL ODE were transferred into a 25-mL three-necked flask and degassed under vacuum for 30 min at 120°C. Then the mixture was placed under N$_2$ for 10 min. Afterwards, 1 mL CsSrCl$_3$:Eu seed solution and 1 mL S-ODE were quickly added in turn and kept for 30 min. The solution was cooled to room temperature by using an ice-water bath. For the purification of ZnS-LNCs, the crude solution was centrifuged at 9000 rpm for 5 min. The supernatant was discarded, while 3 mL of hexane and 3 mL of MeOAc were added to the precipitates and centrifuged again at 7000 rpm for 5 min. The final product was redispersed in 3 mL of hexane for further tests.

**Characterization**

XRD experiments were performed by an X-ray polycrystalline diffractometer (Bruker D2 PHASER) with Cu (Kα) radiation (λ = 1.54 Å). The LNCs samples were prepared in a glove box by drop-casting on a glass substrate, and were subsequently encapsulated with Kapton tape to prevent exposure to moisture and oxygen. TEM images were taken with a Tecnai G2 F20 S-Twin transmission electron microscope at an accelerating voltage of 200 kV. EDS elemental mapping images were recorded using a Tecnai G2 F20 S-Twin microscope. The relative ratio of elements in LNCs was measured by the ICP-MS (Agilent 7800). XPS analysis was measured by using a Thermo Scientific K-Alpha X-ray photoelectron spectrometer. FTIR data was collected on a Theromo Fisher Scientific Nicolet iS20 Fourier transform infrared spectrometer in the transmittance mode. The PL spectra of LNCs was recorded by using a 365 nm UV lamp (6 W) and Ocean View's QE Pro spectrometer. PLE spectra were characterized by fluorescence spectrophotometer (F97XP, Shanghai Lengguang, China). The PLQY of LNCs was recorded on a spectrofluorometer (FluoroMax+, HORIBA) using a Bluefield optics integrating sphere, excited by a 360 nm UV exciter. The PL lifetime and temperature-dependent PL spectra of LNCs was measured by using a steady-state transient florescence spectrometer (Edinburgh FLS1000), excited by a 375 nm exciter. Prior to the measurement of PL lifetime and temperature-dependent PL spectra, LNCs were sandwiched between two pieces of square quartz plates (2 cm × 2 cm), with the edges sealed by UV-curable adhesive in the glove box. The water contact angle of LNCs was measured by using Dataphysics OCA 20, and the samples were prepared in the glove box by spin-coating LNCs solutions on glass substrates. The stability of LNCs was tested in the ambient environment (~25°C, ~40% humidity) under continuous irradiation from a 365 nm UV lamp with a power density of 20 mW/cm$^2$ on the surface, and the samples were prepared in the glove box by drop-casting LNCs solutions on glass substrates.

# Acknowledgements

This work is supported by the Science and Technology Commission of Shanghai Municipality (21ZR1408800).

# Author contributions

**Chuangchang Lei**: Conceptualization, Methodology, Software, Validation, Formal analysis, Investigation, Visualization, Writing – original draft. **Xiang Wu**: Conceptualization, Methodology, Formal analysis, Investigation. **Yaohua Li**: Investigation. **Xu Xu**: Resources. **Guangzheng Zuo**: Resources. **Qiongrong Ou**: Supervision, Resources, Funding acquisition. **Shuyu Zhang**: Conceptualization, Supervision, Resources, Project administration, Funding acquisition, Writing – review & editing. All authors discussed the results and commented on the paper.

# Competing interests

The authors declare no competing interests.

# Data availability

The data that support the findings of this paper are available from the corresponding author upon request.

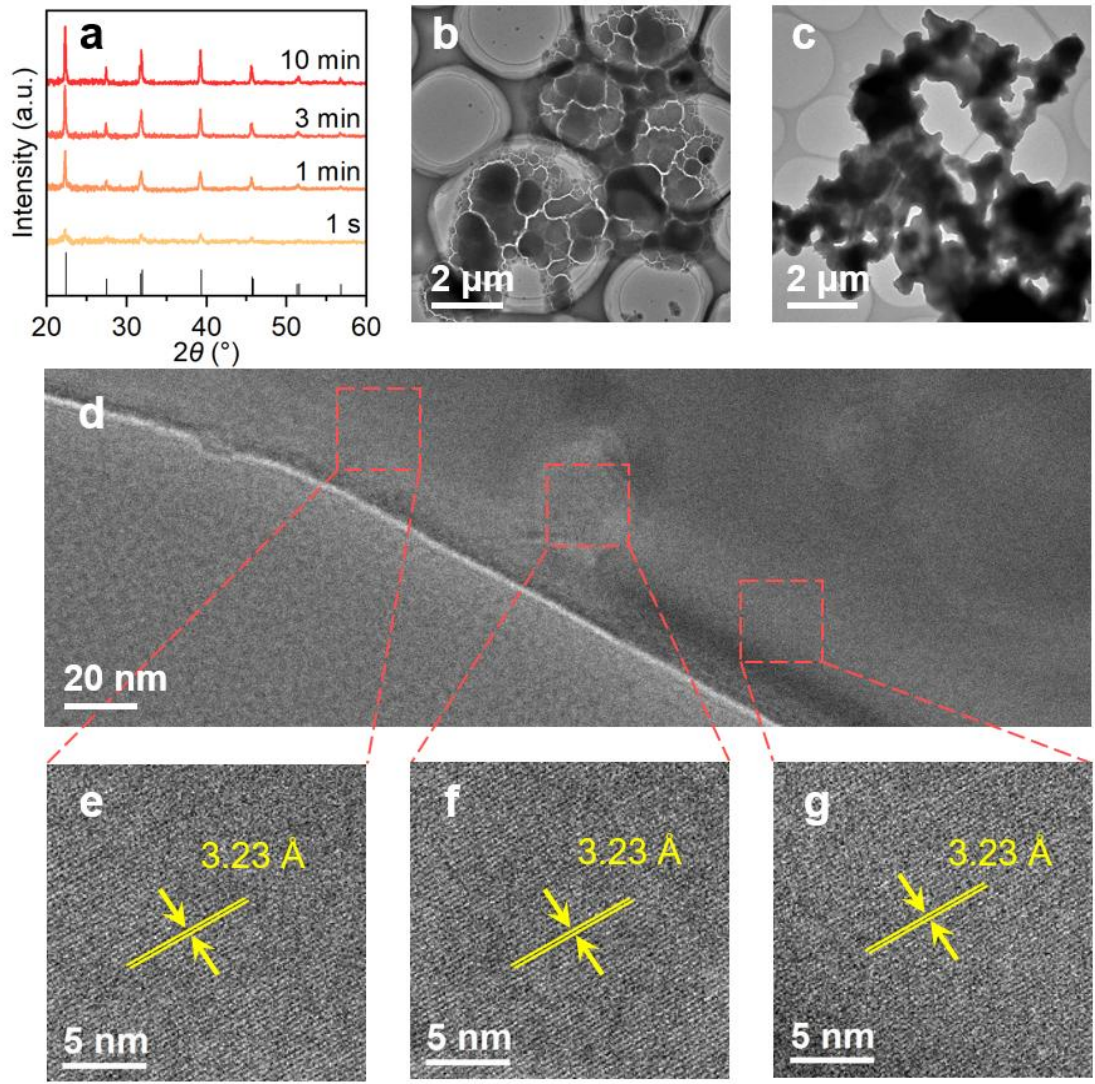

**Fig. 1 Synthesis and characteristics of CsSrCl$_3$:Eu large nanocrystal clusters. a** XRD patterns of samples with reaction time ranging from 1 s to 10 min. The bars at the bottom represent the standard pattern of CsSrCl$_3$ (No. 00-020-0289). TEM images of **b** LNCs-1s and **c** LNCs-10min. **d-g** HRTEM images of LNCs-10min. **e**, **f** and **g** are the magnified images of selected regions in **d**.

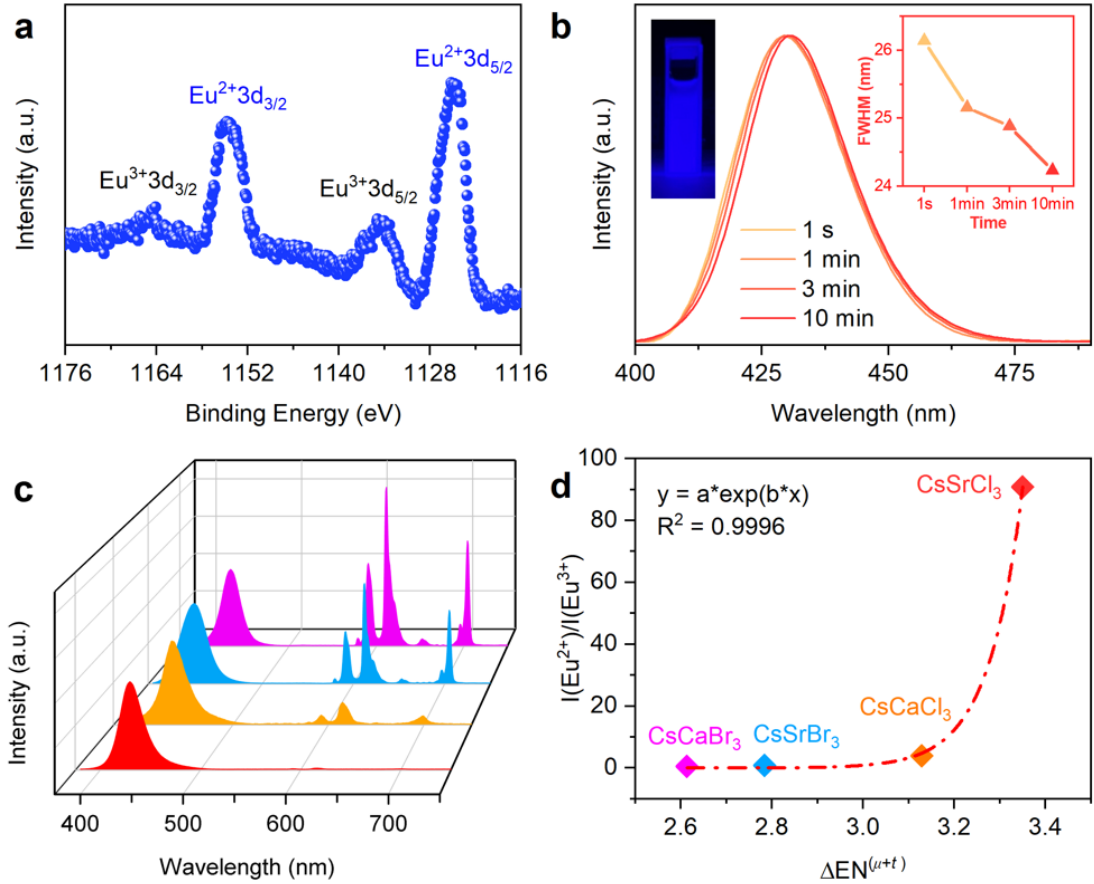

**Fig. 2 Self-reduction effect of Eu$^{3+}$ to Eu$^{2+}$. a** X-ray photoelectron spectroscopy (XPS) analysis of Eu in LNCs-10min. **b** PL spectra of CsSrCl$_3$:Eu samples with reaction time from 1s to 10 min, excited by a 365 nm UV lamp. Insets show the change of FWHM as a function of reaction time and the picture of LNCs under UV lamp. **c** PL spectra of CsSrCl$_3$:Eu (red), CsCaCl$_3$:Eu (orange), CsSrBr$_3$:Eu (blue) and CsCaBr$_3$:Eu (magenta), excited by a 275 nm lamp. **d** The PL intensity ratio of I(Eu$^{2+}$)/I(Eu$^{3+}$) against △EN$^{(\mu+t)}$ in CsSrCl$_3$:Eu, CsCaCl$_3$:Eu, CsSrBr$_3$:Eu and CsCaBr$_3$:Eu, along with corresponding fitted curve.

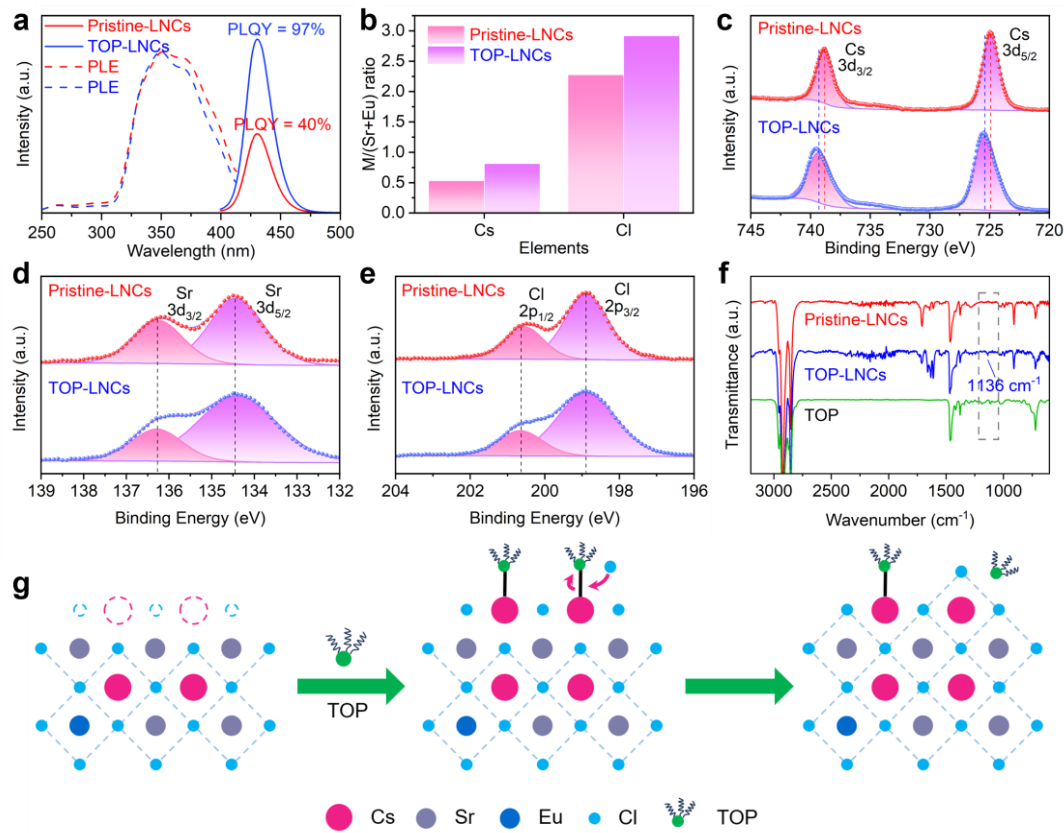

**Fig. 3 The surface passivation mechanism of TOP post-synthetic treatment. a** PL and PLE spectra of Pristine-LNCs and TOP-LNCs. **b** Relative element ratio of M/(Sr+Eu) (M = Cs, Cl) in Pristine-LNCs and TOP-LNCs from the quantitative XPS analysis. XPS spectra of **c** Cs, **d** Sr and **e** Cl in Pristine-LNCs and TOP-LNCs. **f** FTIR spectra of Pristine-LNCs, TOP-LNCs and TOP. **g** Illustration of surface treatment mechanism by adding TOP.

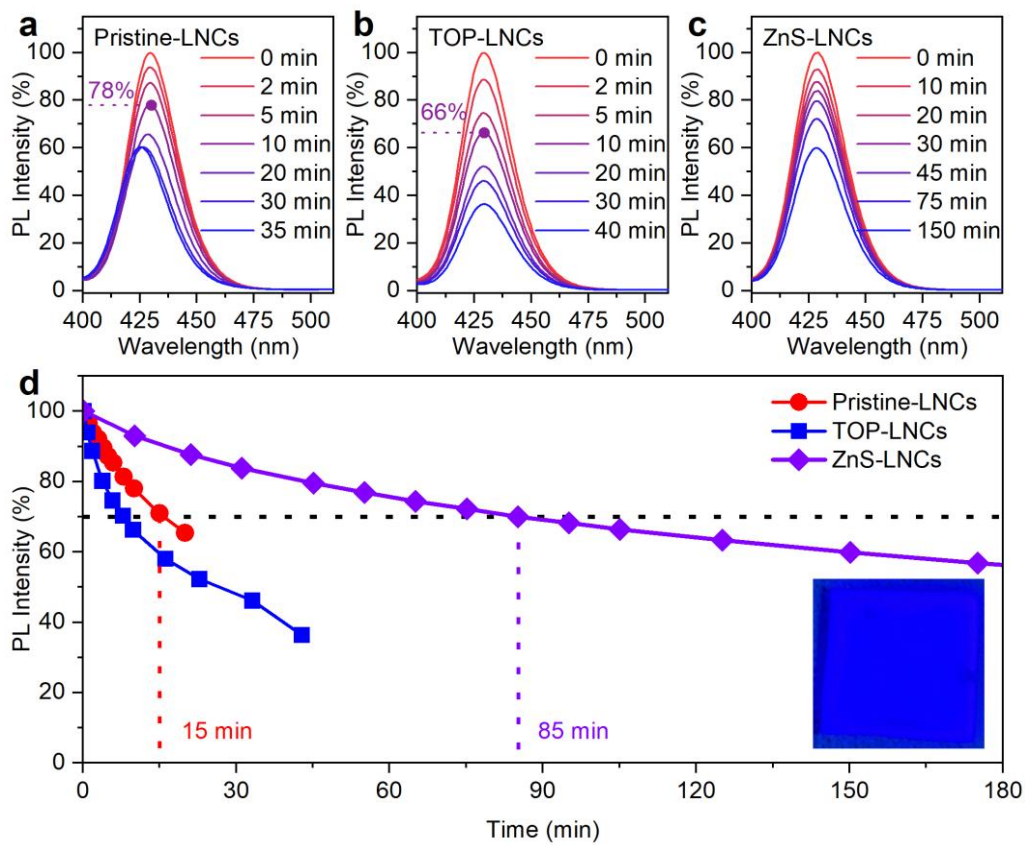

**Fig. 4 Improving the stability of CsSrCl₃:Eu by growing a shell of ZnS.** Changes in PL spectra of thin films made from **a** Pristine-LNCs, **b** TOP-LNCs and **c** ZnS-LNCs in ambient environment under continuous irradiation, excited by a 365 nm UV light. **d** Recordings of PL peak intensity of Pristine-LNCs, TOP-LNCs and ZnS-LNCs during the continuous irradiation. Pristine-LNCs was recorded for only 20 min since the peak position was shifted obviously at 30 min, indicating a decomposition of perovskite structure. Inset shows the picture of the LNCs-film under UV lamp.

Supplementary Information for

# Eu-doped CsSrCl$_3$ Large Nanocrystal Clusters with Self-Reduction Effect and Near-Unity Quantum Yield

Chuangchang Lei[1], Xiang Wu[1], Yaohua Li[1], Xu Xu[2], Guangzheng Zuo[1], Qiongrong Ou[1], Shuyu Zhang[1,*]

[1]State Key Laboratory of Photovoltaic Science & Technology, School of Intelligent Robotics and Advanced Manufacturing, Institute for Electric Light Sources, Fudan University, Shanghai 200433, PR China

[2] Donghai Laboratory, Zhoushan, 316021, PR China

E-mail: forxuxu@126.com; gzzuo@fudan.edu.cn

*Corresponding author: zhangshuyu@fudan.edu.cn

**Supplementary Note 1:**

The structures of the molecules are optimized at B97-3c level[1] using ORCA 5.0.4 software[2]. Using B3LYP-D3(BJ)/def2-TZVP[3–5] in Gaussian 16 A.03 software, the binding energy was calculated by

$$\Delta E = E_{AB} - E_A - E_B + E_{BSSE} \tag{S1}$$

where $E_{AB}$ is the energy of the complex, $E_A$ or $E_B$ is the energy of the monomer, and $E_{BSSE}$ is the basis set superposition error (BSSE) correction energy obtained using the counterpoise method[6]. The structures are visualized using VMD 1.9.3[7] software.

**Supplementary Note 2:**

PLQY can be defined as the proportion of radiative recombination rate in total recombination rate, which is:

$$\text{PLQY} = \frac{k_r}{k_r + k_{nr}} \tag{S2}$$

where $k_r$ is the radiative recombination rate and $k_{nr}$ is the nonradiative recombination rate.

The average PL lifetime refers to the average time that excitons stay in the excited state after photoexcitation before they return to the ground state through radiative or nonradiative pathways. The average PL lifetime is calculated as:

$$\tau_{ave} = \frac{1}{k_r + k_{nr}} = \frac{A_1 \tau_1^2 + A_2 \tau_2^2}{A_1 \tau_1 + A_2 \tau_2} \tag{S3}$$

where $A_1$ and $A_2$ are constants, $\tau_1$ and $\tau_2$ are decay times.

Then, the radiative recombination rate and nonradiative recombination rate can be calculated as:

$$k_r = \frac{\text{PLQY}}{\tau_{ave}} \tag{S4}$$

$$k_{nr} = \frac{1}{\tau_{ave}} - k_r = \frac{1-\text{PLQY}}{\tau_{ave}} \tag{S5}$$

**Supplementary Note 3:**

Then integrated temperature-dependent PL intensity of TOP-LNCs was fitted by:

$$I(T) = \frac{I_0}{1+A\exp(-\frac{E_b}{k_B T})} \tag{S6}$$

where $I(T)$ is the integrated PL intensity at different temperature, $I_0$ is the integrated PL intensity at 0 K, A is the radiative decay constant, $E_b$ is exciton binding energy and $k_B$ is the Boltzmann's constant.

Huang-Rhys factor ($S$) is used to determine the strength of electron-phonon coupling, which can be obtained by fitting the temperature-dependent FWHM against temperature:

$$\text{FWHM} = 2.36\sqrt{S}\hbar\omega_{phonon}\sqrt{\coth\frac{\hbar\omega_{phonon}}{2k_B T}} \tag{S7}$$

where $\hbar$ is the reduced Plank constant and $\omega_{phonon}$ is the phonon frequency.

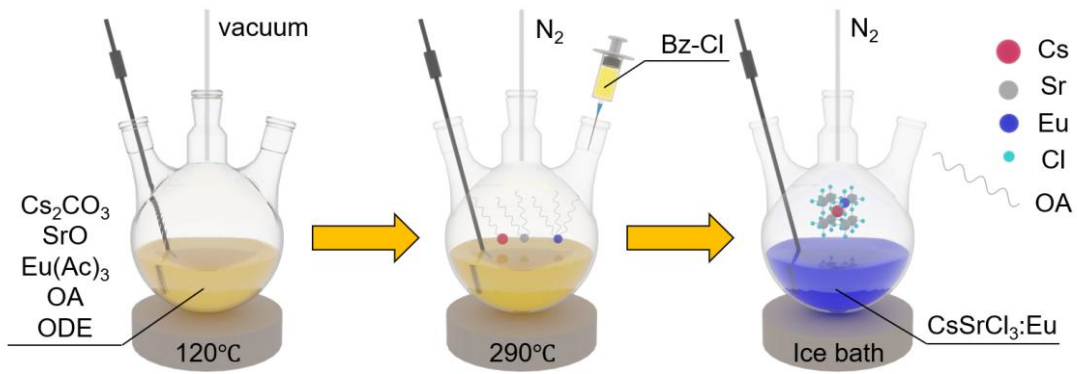

**Supplementary Fig. 1 | Schematic diagram of the one-step-synthesis for CsSrCl₃:Eu.**

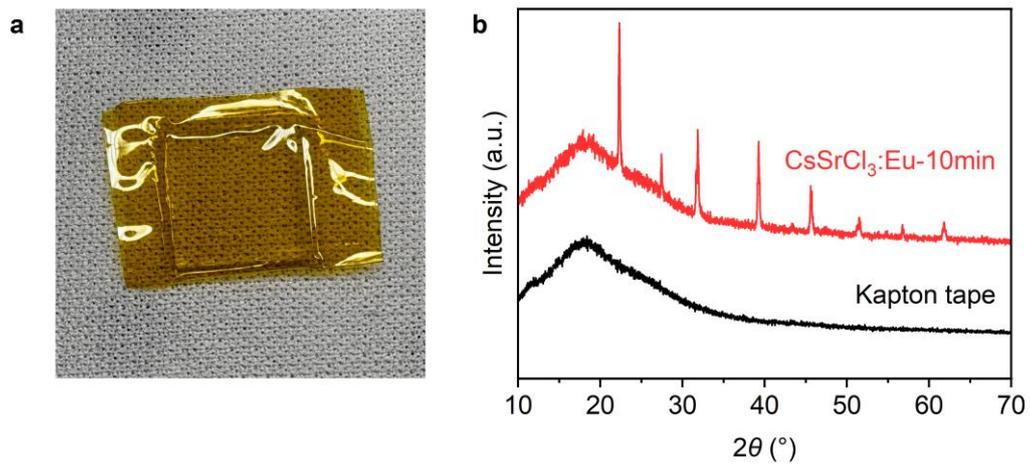

**Supplementary Fig. 2 | Kapton tape's XRD spectra.** (a) Blank glass substrate encapsulated by Kapton tape. (b) X-ray diffraction (XRD) spectra of Kapton tape and CsSrCl$_3$:Eu (reacted for 10 min).

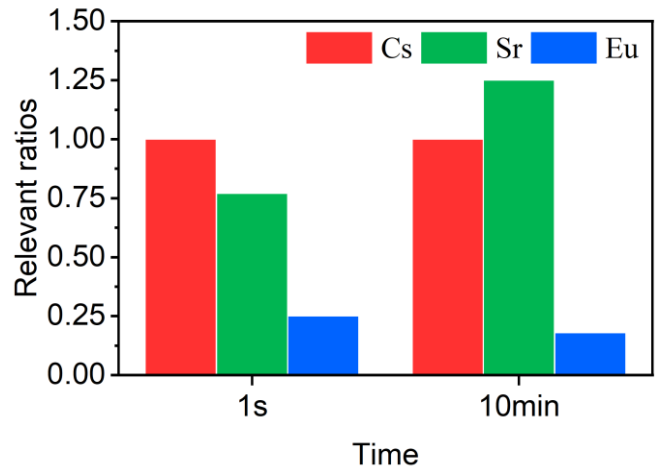

**Supplementary Fig. 3 | Inductively coupled plasma mass spectrometry (ICP-MS) analyses of LNCs-1s and LNCs-10min.**

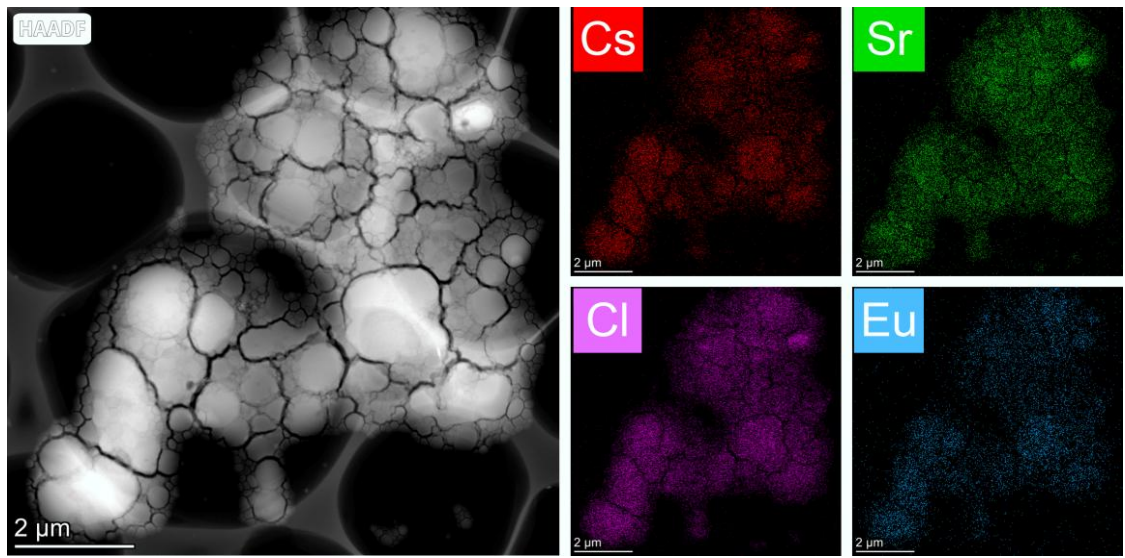

**Supplementary Fig. 4 | Energy-dispersive spectroscopy (EDS) mapping of LNCs-1s.**

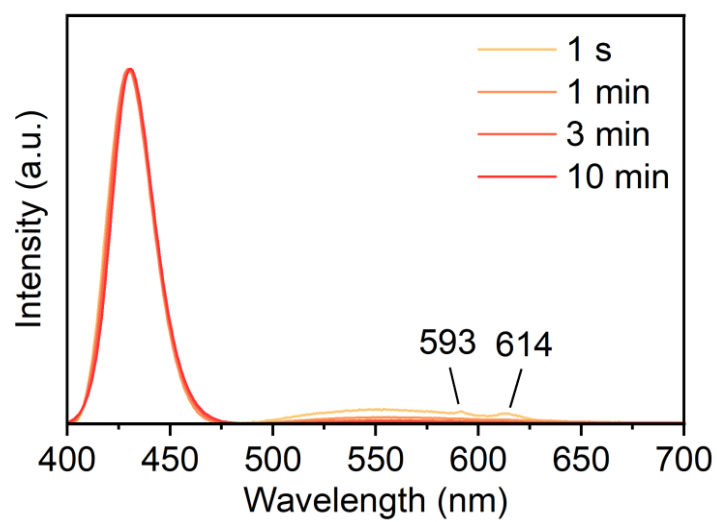

**Supplementary Fig. 5 | Normalized PL spectra of LNCs from 1 s to 10 min within visible spectrum.** The characteristic peaks of $Eu^{3+}$ (593 nm and 614 nm) in LNCs-1s originate from the $Eu^{3+}$ ions that have not yet been completely reduced in the early stage of the reaction, which would disappear quickly once the reaction time is extended.

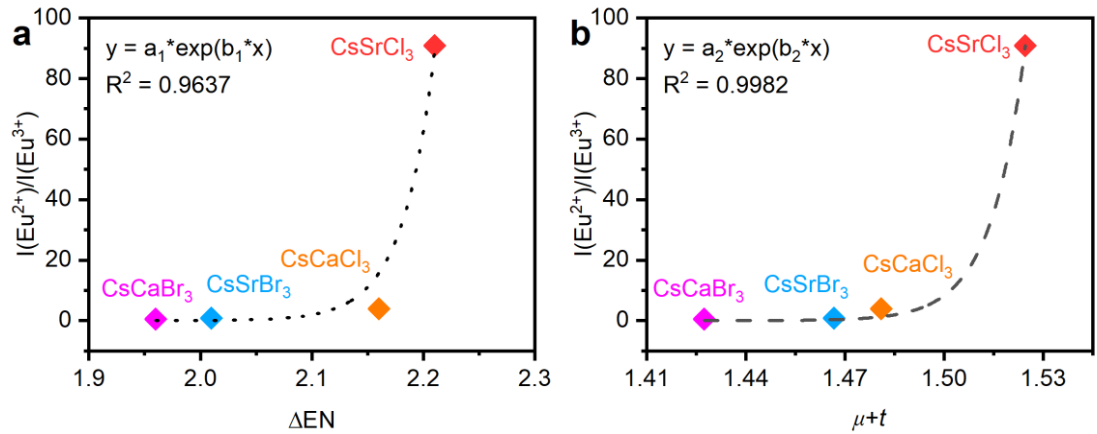

**Supplementary Fig. 6 | PL intensity ratio of I(Eu$^{2+}$)/I(Eu$^{3+}$) against $\triangle$EN and ($\mu$+$t$) in AEM-based halide perovskites.** The PL intensity ratio of I(Eu$^{2+}$)/I(Eu$^{3+}$) against **a** $\triangle$EN between AEM and halogens, **b** ($\mu$+$t$) in CsSrCl$_3$:Eu, CsCaCl$_3$:Eu, CsSrBr$_3$:Eu and CsCaBr$_3$:Eu, along with corresponding fitted curves.

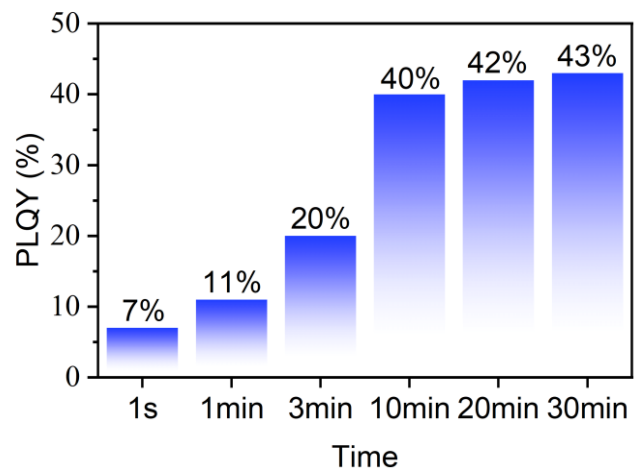

**Supplementary Fig. 7 | PLQY values of LNCs with reaction time from 1 s to 30 min.**

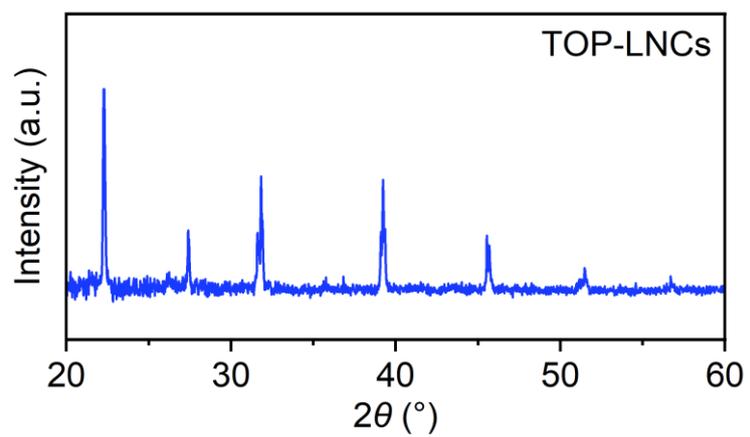

**Supplementary Fig. 8 | XRD pattern of TOP-LNCs.**

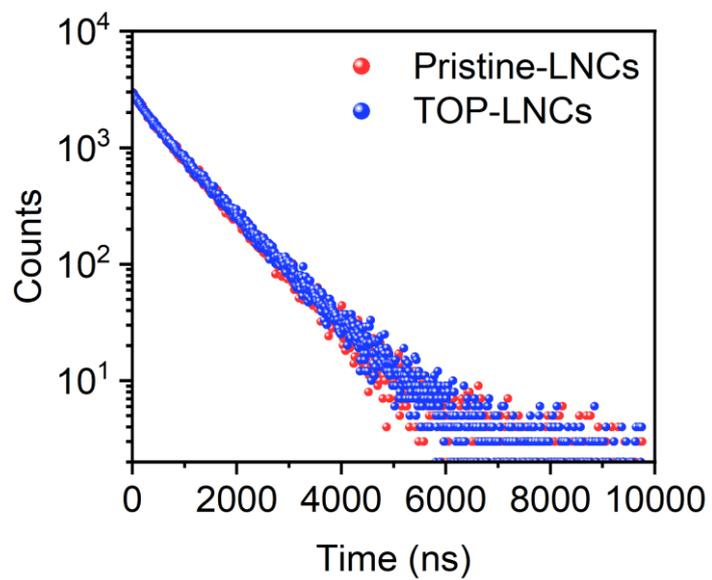

**Supplementary Fig. 9 | Time-resolved photoluminescence (TRPL) spectra of Pristine-LNCs and TOP-LNCs.**

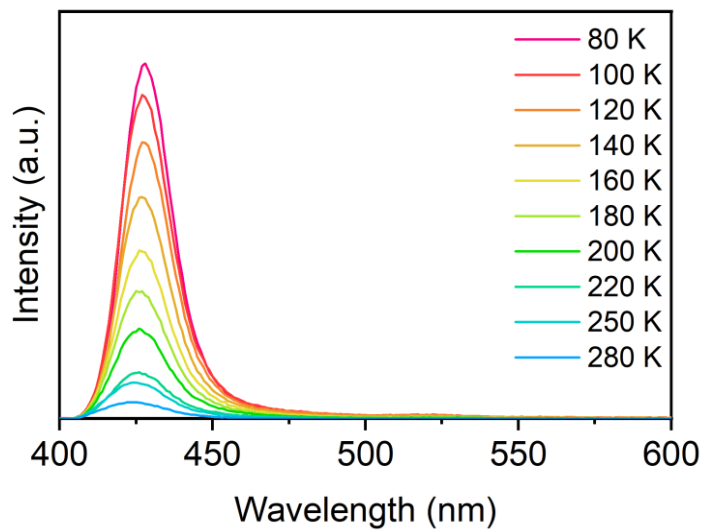

**Supplementary Fig. 10 | Temperature-dependent PL spectra of TOP-LNCs.** Temperature-dependent PL spectra of TOP-LNCs between 80 and 280 K, excited by a 375 nm light.

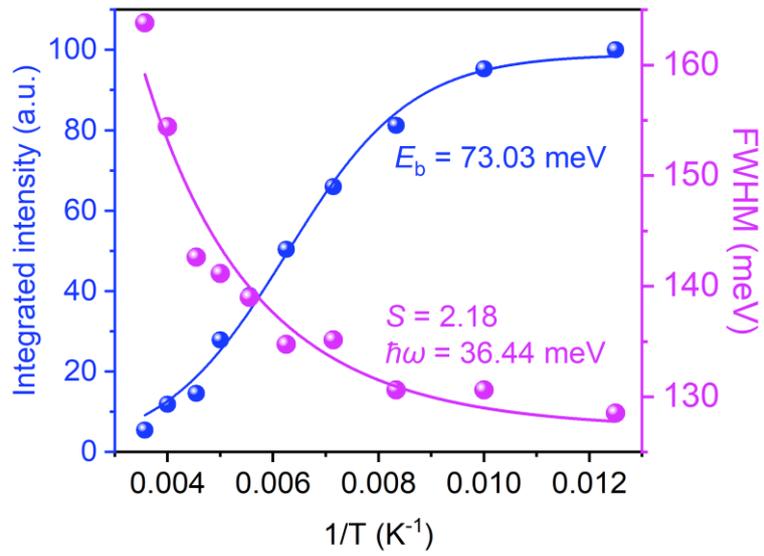

**Supplementary Fig. 11 | E$_b$ and $\hbar\omega_{phonon}$ of TOP-LNCs.** Integrated PL intensity and FWHM of TOP-LNCs as a function of temperature.

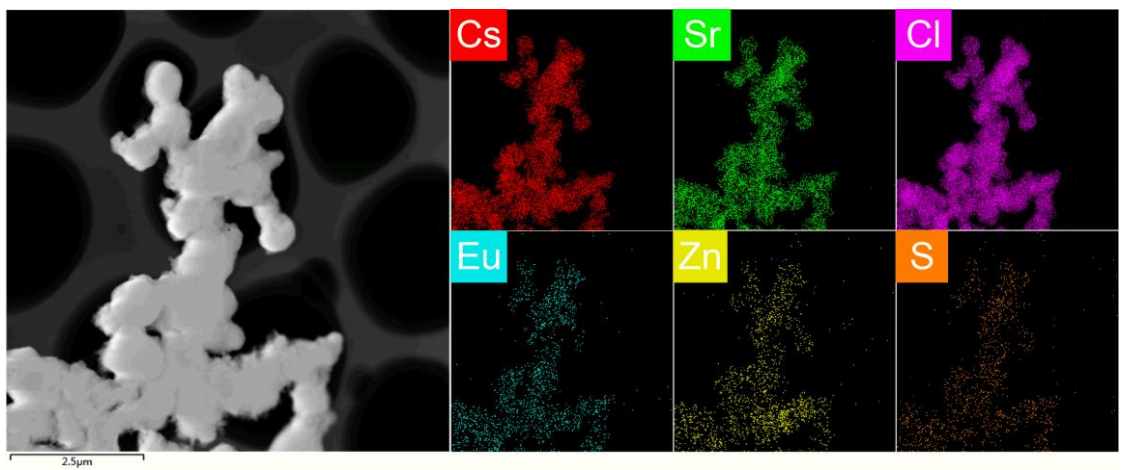

**Supplementary Fig. 12 | EDS elemental mapping of ZnS-LNCs.**

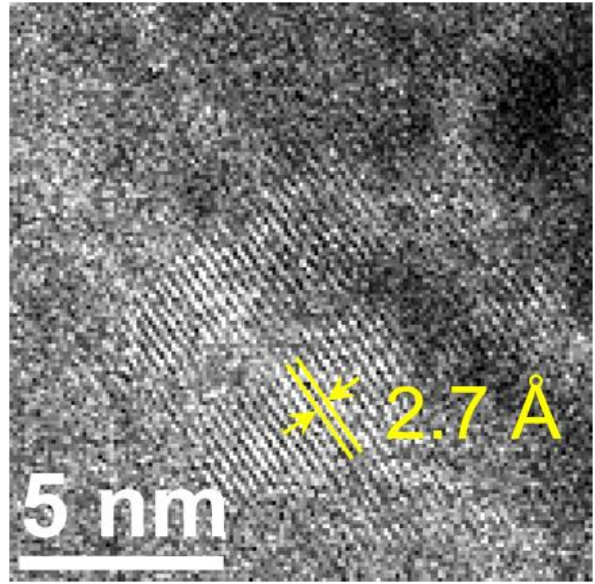

**Supplementary Fig. 13 | TEM image of ZnS-LNCs.**

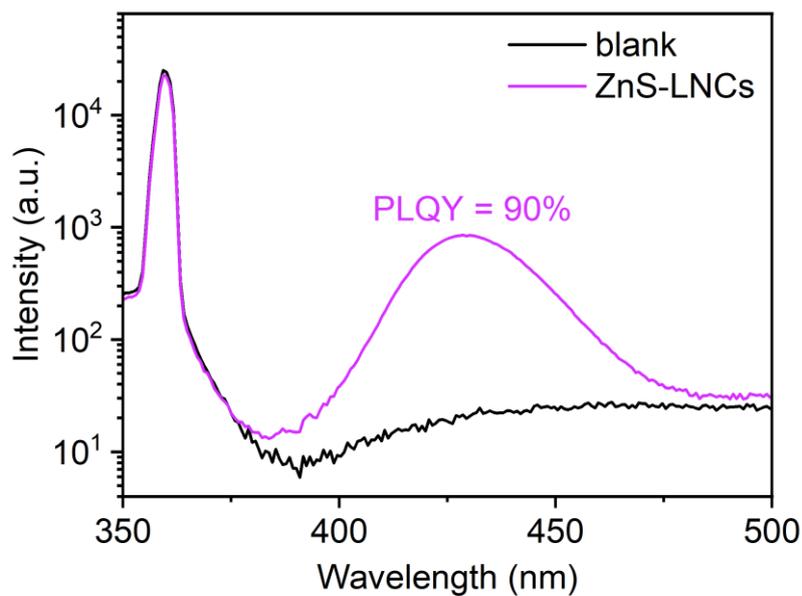

**Supplementary Fig. 14 | Photoluminescence quantum yields (PLQY) spectra of ZnS-LNCs.**

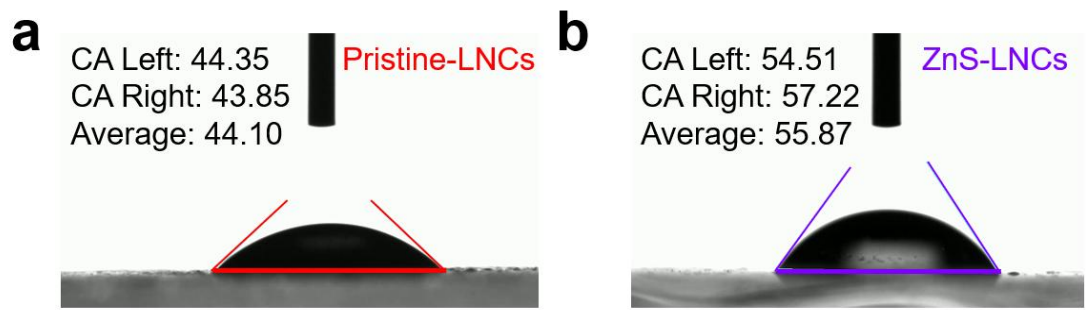

**Supplementary Fig. 15 | Water contact angle (CA) of (a) Pristine-LNCs and (b) ZnS-LNCs.**

**Supplementary Table 1 | The values of tolerance factor (*t*), octahedral factor (*μ*) and electronegativity (△EN) in AEM-based halide perovskites.**

| Perovskites | $t$ | $\mu$ | $\mu+t$ | $\triangle EN$ | $\triangle EN^{(\mu+t)}$ |
|---|---|---|---|---|---|
| $CsCaBr_3$ | 0.510 | 0.917 | 1.427 | 1.96 | 2.613 |
| $CsSrBr_3$ | 0.602 | 0.865 | 1.467 | 2.01 | 2.784 |
| $CsCaCl_3$ | 0.552 | 0.929 | 1.481 | 2.16 | 3.129 |
| $CsSrCl_3$ | 0.652 | 0.873 | 1.525 | 2.21 | 3.350 |

**Supplementary Table 2 | The values of radiative recombination ($k_r$) and nonradiative recombination ($k_{nr}$) rates in Pristine-LNCs and TOP-LNCs.**

| Samples | $A_1$ | $\tau_1$ (ns) | $A_2$ | $\tau_2$ (ns) | $\tau_{ave}$ (ns) | PLQY (%) | $k_r$ (μs$^{-1}$) | $k_{nr}$ (μs$^{-1}$) |
|---|---|---|---|---|---|---|---|---|
| Pristine-LNCs | 592.3 | 275 | 2932.7 | 880 | 836 | 40 | 0.514 | 0.682 |
| TOP-LNCs | 610.4 | 269 | 2368.2 | 899 | 854 | 97 | 1.113 | 0.035 |

**Supplementary References:**